\documentclass[aps,prb,twocolumn,amsfonts,superscriptaddress,showpacs,final,floatfix,preprintnumbers]{revtex4-1}

\usepackage{amsmath, amssymb, mathrsfs}
\usepackage{latexsym}
\usepackage{graphicx,epstopdf,color}
\usepackage{braket}
\usepackage{subfigure}
\usepackage{times}
\usepackage{bm}

\usepackage[breaklinks]{hyperref}
\hypersetup{
   colorlinks,
   citecolor=blue,
   filecolor=blue,
   linkcolor=blue,
   urlcolor=blue
}

\newcommand{\hc}{\textrm{h.c.}}
\newcommand{\mf}[1]{\mathfrak{#1}}
\newcommand{\mc}[1]{\mathcal{#1}}
\newcommand{\mb}[1]{\mathbb{#1}}
\renewcommand{\vec}[1]{\bm{#1}}

\allowdisplaybreaks[1]

\begin{document}

\title{Emergent Chiral Spin State in the Mott Phase of a Bosonic Kane-Mele-Hubbard Model}
\author{Kirill Plekhanov}
\affiliation{LPTMS, CNRS, Univ. Paris-Sud, Universit\'e Paris-Saclay, 91405 Orsay, France}
\affiliation{Centre de Physique Th\'eorique, Ecole Polytechnique,
  CNRS, Universit\'e Paris-Saclay, F-91128 Palaiseau, France}
\author{Ivana Vasi\'c}
\affiliation{
  Scientific Computing Laboratory, Center for the Study of Complex
Systems, Institute of Physics Belgrade, University of Belgrade, 11080 Belgrade, Serbia
}
\author{Alexandru Petrescu}
\affiliation{Department of Electrical Engineering, Princeton
  University, Princeton, New Jersey, 08544}
\author{Rajbir Nirwan}
\affiliation{Institut f\"{u}r Theoretische Physik, Goethe-Universit\"{a}t,
  60438 Frankfurt/Main, Germany}
\author{Guillaume Roux}
\affiliation{LPTMS, CNRS, Univ. Paris-Sud, Universit\'e Paris-Saclay,
  91405 Orsay, France}
\author{Walter Hofstetter}
\affiliation{Institut f\"{u}r Theoretische Physik, Goethe-Universit\"{a}t,
  60438 Frankfurt/Main, Germany}
\author{Karyn Le Hur}
\affiliation{Centre de Physique Th\'eorique, Ecole Polytechnique, CNRS, Universit\'e Paris-Saclay, F-91128 Palaiseau, France}
\date{\today}

\begin{abstract}
  Recently, the frustrated XY model for spins-1/2 on the honeycomb
  lattice has attracted a lot of attention in relation with the
  possibility to realize a chiral spin liquid state. This model is
  relevant to the physics of some quantum magnets. Using the
  flexibility of ultra-cold atoms setups, we propose an alternative
  way to realize this model through the Mott regime of the bosonic
  Kane-Mele-Hubbard model.  The phase diagram of this model is derived
  using the bosonic dynamical mean-field theory. Focussing on the Mott
  phase, we investigate its magnetic properties as a
  function of frustration. We do find an emergent chiral spin
  state in the intermediate frustration regime. 
  Using exact diagonalization we study more closely the physics
  of the effective frustrated XY model and the properties of the chiral spin state. This gapped phase
  displays a chiral order, breaking time-reversal and parity symmetry,
  but is not topologically ordered (the Chern number is zero).
\end{abstract}


\maketitle

The last few decades have seen a growing interest in the quest for
exotic spin states and quantum spin liquids~\cite{Lhuillier2002}.
Significant progress has been made both from the theoretical and
experimental sides~\cite{Balents2010, Norman2016, SavaryBalents2017}.
The best candidates for spin liquids are found in two-dimensional
systems.  Disordered phases are expected to occur in complex geometries,
such as the Kagome lattice~\cite{Lecheminant1997, YanHuseWhite2011,
  Depenbrock2012}, or in frustrated bipartite lattices, such as the
square lattice with second-neighbor couplings~\cite{SchulzZiman1992,
  Schulz1996}. Among basic lattices, the honeycomb one hosts free
Majorana fermions due to Kitaev anisotropic
interactions~\cite{Kitaev2006}, and raises questions when starting
from the Hubbard model~\cite{Meng2010,Sorella2012,Assaad2013}.  In
such context and motivated by quantum magnets~\cite{Flint2013},
frustrated Heisenberg models on the honeycomb lattice have been
recently explored~\cite{FouetSindzingreLhuillier2001, Wang2010,
  MudlerEtAl2010, Clark2011, Albuquerque2011, Cabra2011, Reuther2011,
  Mezzacapo2012, Zhang2013, Ganesh2013, GongShengEtAl2013, Zhu2013,
  GonfZhuBalentsSheng2015, Ferrari2017}.  In parallel, the XY version
of this model was also tested for the possibility to realize a chiral
spin liquid state, but with seemingly contradictory
results~\cite{VarneySunGalitskiRigol2011, VarneySunGalitskiRigol2012,
  ZhuHuseWhite2013, ZhuWhite2014, BishopLiCampbell2014,
  CarrasquillaEtAl2013, CioloEtAl2014, NakafujiIchinose2017}.  As
suggested in Ref.~\onlinecite{SedrakyanGlazmanKamenev2015}, in the
intermediate frustration regime the ground-state physics could be
mapped to a fermionic Haldane model~\cite{Haldane1988} with
topological Bloch bands at a mean-field level, as a result of
Chern-Simons (ChS) gauge fields~\cite{Fradkin1989,
  AnbjornSemenoff1989, LopezRojoFradkin1994,
  Misguichjolicoeurgirvin2001, SunKumarFradkin2015}. However, the
topological nature of this spin state is still elusive.

Our objectives are two-fold in this Letter. Motivated by cold atoms
experiments~\cite{BlochDalibardNascimbene2012,
  GoldmanBudichZoller2016}, we first show that the Mott regime of the
bosonic Kane-Mele-Hubbard (BKMH) model allows for a tunable
realization of the frustrated XY model on the honeycomb lattice.
Second, we study its phase diagram and in particular its magnetic
properties, using bosonic dynamical mean-field theory
(B-DMFT)~\cite{GeorgesKotliarKrauthRozenberg1996, ByczukVollhardt2008,
  HuTong2009, HubenerSnoekHofstetter2009,
  AndersGullPolletTroyerWerner2010, suppMaterial}, exact
diagonalization (ED) and theoretical arguments. The Kane-Mele
model~\cite{KaneMele2005} is the standard model with spin-orbit
coupling that displays $\mb{Z}_2$ topology. Still, it has not yet been
studied for interacting bosons. Importantly, we recall that, for
interacting fermions and at the Mott transition, the Kane-Mele model
becomes magnetically ordered in the $xy$-plane, with quantum
fluctuations stabilizing the N\'eel ordering~\cite{RachelLehur2010,
  WuRachelLiuLehur2012, Hohenadler2012}.

\begin{figure}
  \includegraphics[width=0.39\textwidth] {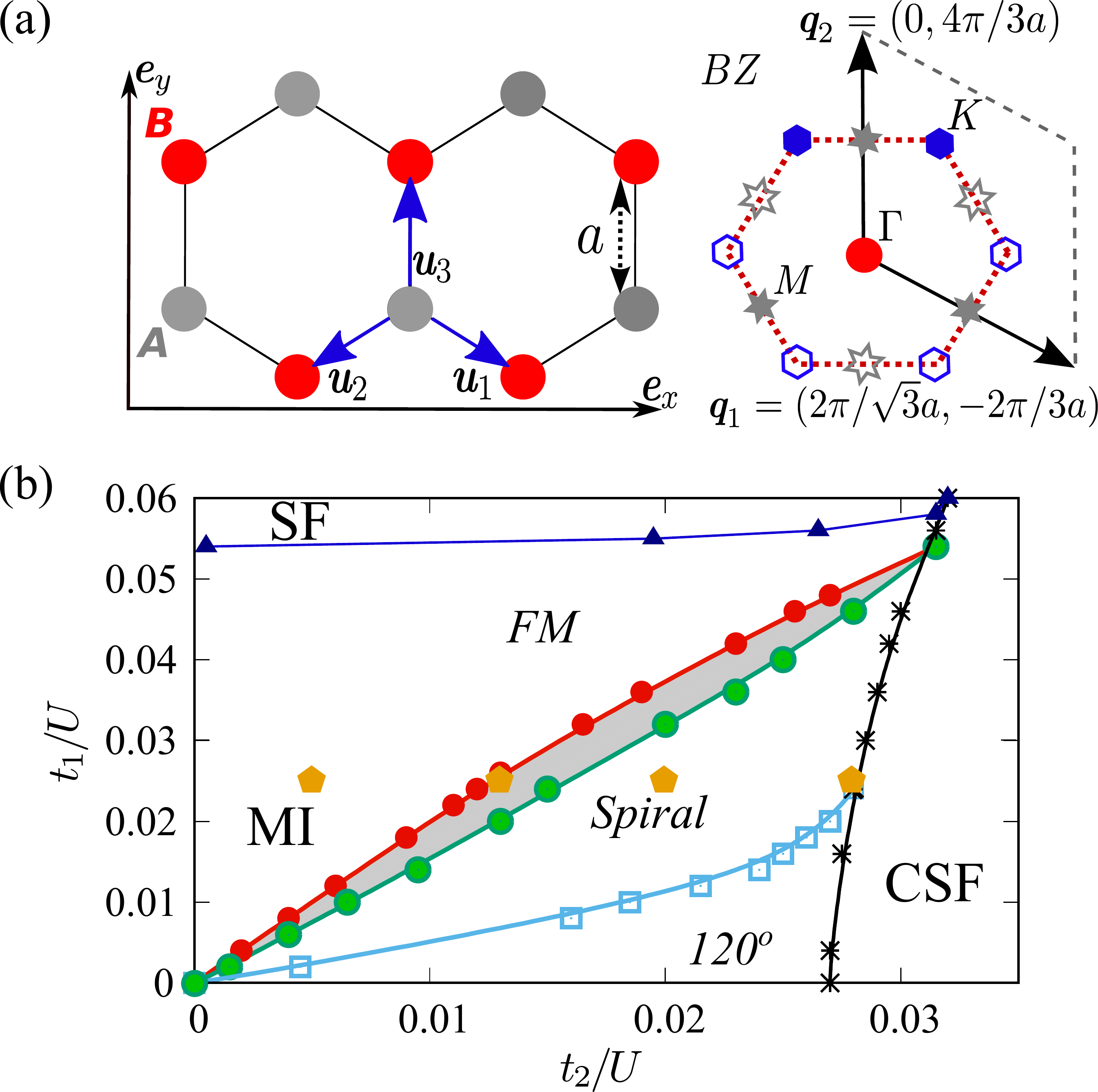}
  \caption{\textbf{(a)} Honeycomb lattice with $\vec{u}_i$ -- vectors
    between first neighbor sites and the first Brillouin zone with
    explicitly shown $\vec{\Gamma}$, $\vec{K}$ and $\vec{M}$
    points. \textbf{(b)} Phase diagram of the BKMH model obtained
    using B-DMFT containing Mott insulator (MI), uniform superfluid
    (SF) and chiral superfluid (CSF) phases with different regimes of
    the MI phase marked in italic. The central gray region corresponds
    to the states with no coplanar order. Parameters
    $U_{\uparrow \downarrow}/U = 0.5,\ \mu/U_{\uparrow \downarrow} =
    0.5$, lattice of 96 sites. "Pentagons" mark parameter values that
    we further explore in Fig.~\ref{fig:1}(a-d).}
  \label{fig:0}
\end{figure}

We start our analysis with the bosonic version of the Kane-Mele
model~\cite{KaneMele2005} on the honeycomb lattice
(Fig.~\ref{fig:0}(a)), which contains two species of bosons labelled
by $\sigma = \uparrow, \downarrow$. In the presence of Bose-Hubbard
interactions, the Hamiltonian reads:
\begin{align}
 \label{eq:bkmhModel}
  H =
  & - t_1 \!\!\sum\limits_{\sigma, \Braket{ij}} [b^\dag_{\sigma,\vec{r}_i} b_{\sigma,\vec{r}_j} + \hc ]
    + i t_2 \!\!\!\!\sum\limits_{\sigma, \Braket{\Braket{ik}}}
    \!\!\!\! \nu^\sigma_{ik} [ b^\dag_{\sigma,\vec{r}_i}
    b_{\sigma,\vec{r}_k} - \hc ]
    \notag \\
  & + \frac{U}{2} \sum\limits_{\sigma,i} n_{\sigma, \vec{r}_i}
    (n_{\sigma, \vec{r}_i} - 1) + U_{\uparrow \downarrow}
    \sum\limits_{i} n_{\uparrow, \vec{r}_i} n_{\downarrow, \vec{r}_i}
    \;.
\end{align}
Here, $b^\dag_{\sigma,\vec{r}_i} (b_{\sigma,\vec{r}_i})$ are creation
(annihilation) operators at site $i$ of the honeycomb lattice, and
$n_{\sigma,\vec{r}_i} = b^\dag_{\sigma,\vec{r}_i}
b_{\sigma,\vec{r}_i}$ is the density operator. $t_1\ \text{(resp. } t_2$) is
the amplitude of hopping to the first (resp. second) neighbors and
$\nu^\uparrow_{ik} = -\nu^\downarrow_{ik} = 1\ (\text{resp. } -1)$ for hoppings
corresponding to a left-turn (resp. right-turn) on the honeycomb lattice.  We assume a filling
of one boson per site
$\Braket{n_{\uparrow,\vec{r}_i} + n_{\downarrow,\vec{r}_i}} = 1$.  The
Haldane model~\cite{Haldane1988} for spinless fermions has been
realized through Floquet engineering in cold atoms~\cite{Jotzu2014}.
Similarly, spin-orbit models have been proposed in optical lattices
setups~\cite{KennedyEtAl2013, StruckEtAl2013, Yan2015} and
experimentally achieved with photons~\cite{HafeziEtAl2011,
  SalaEtAl2015, LuJoannopoulosSoljacic2014, LeHurEtAl2016}.  All the
ingredients required for a successful implementation of
\eqref{eq:bkmhModel} are thus available.

\textit{I. B-DMFT on BKMH model.} The ground-state phase diagram of the BKMH model obtained from
B-DMFT~\cite{GeorgesKotliarKrauthRozenberg1996, ByczukVollhardt2008,
  HuTong2009, HubenerSnoekHofstetter2009,
  AndersGullPolletTroyerWerner2010} is shown in
Fig.~\ref{fig:0}(b). In order to address unusual states that break
translational symmetry, we use real-space B-DMFT \cite{LiHof, HeLiHof,
  HeJiHof, suppMaterial}. Local effective problems represented by the
Anderson impurity model are solved using exact diagonalization
\cite{suppMaterial}. As found for the bosonic Haldane model with same
filling~\cite{VasicEtAl2015}, three phases are competing: a uniform
superfluid (SF), a chiral superfluid (CSF) and a Mott insulator (MI)
(they are sorted out from the behaviors of the order parameter
$\braket{b_{\sigma,\vec{r}_i}}$ and the local currents
$J^\sigma_{ij} = \mf{Im} \braket{b^\dag_{\sigma,\vec{r}_i}
  b_{\sigma,\vec{r}_j}}$~\cite{suppMaterial}).

We now focus on the MI phase.  As shown in Fig.~\ref{fig:0}(b), the
system enters the Mott phase when intra-species ($U$) and
inter-species ($U_{\uparrow \downarrow}$) interactions become strong
enough.  Applying standard perturbation
theory~\cite{KuklovSvistunov2003}, one rewrites the
Hamiltonian~\eqref{eq:bkmhModel} in terms of pseudo spin-$1/2$
operators
$S^+_{\vec{r}_i} = S^x_{\vec{r}_i} + i S^y_{\vec{r}_i} =
b^\dag_{\uparrow,\vec{r}_i} b_{\downarrow,\vec{r}_i}$,
$S^-_{\vec{r}_i} = S^x_{\vec{r}_i} - i S^y_{\vec{r}_i} =
b^\dag_{\downarrow,\vec{r}_i} b_{\uparrow,\vec{r}_i}$ and
$S^z_{\vec{r}_i} = ( n_{\uparrow,\vec{r}_i} - n_{\downarrow,\vec{r}_i}
) / 2$ as follows:
\begin{align}
  \label{eq:xyModel}
  H =
  & - \sum\limits_{\Braket{ij}} \left[ J_1 \left( S_{\vec{r}_i}^+ S_{\vec{r}_j}^-
    + \hc \right) - K_1 S_{\vec{r}_i}^z S_{\vec{r}_j}^z \right]
    \notag \\
  & + \sum\limits_{\Braket{\Braket{ik}}} \left[ J_2 \left( S_{\vec{r}_i}^+ S_{\vec{r}_k}^-
    + \hc \right) + K_2 S_{\vec{r}_i}^z S_{\vec{r}_k}^z \right]
    \;,
\end{align}
where $J_i = t_i^2 / U_{\uparrow \downarrow}$ and
$K_i = t_i^2 \left( 1/U_{\uparrow \downarrow} - 2/U \right)$.  We
observe that the spin-$1/2$ frustrated XY model is realized when
$U = 2 U_{\uparrow \downarrow}$ (for which $K_i = 0$).  Frustration is
associated with the positive sign of the $J_2$-term, which combines
the sign of the bosonic exchange and the phase of $\pi$ accumulated in
the hoppings between second neighbors.  The fermionic Kane-Mele model
does not include such frustrating terms~\cite{YoungLeeKallin2008,
  RachelLehur2010}. The properties of this effective XY model depend
only on the ratio $ J_2 / J_1 = \left( t_2 / t_1 \right)^2 $.  In the
classical limit, a coplanar ansatz~\cite{RastelliTassiReatto1979,
  FouetSindzingreLhuillier2001, suppMaterial} provides the following
phase diagram: the ferromagnetic phase is stable for
$J_2 / J_1 \leq 1/6$, above which degenerate incommensurate spiral
waves become energetically favoured.  Their wave-vectors leave on
closed contours in the Brillouin zone.  In the case of the Heisenberg
model, quantum fluctuations were predicted to lift this degeneracy via
an order by disorder mechanism~\cite{MudlerEtAl2010}.

\begin{figure}
  \includegraphics[width=0.46\textwidth] {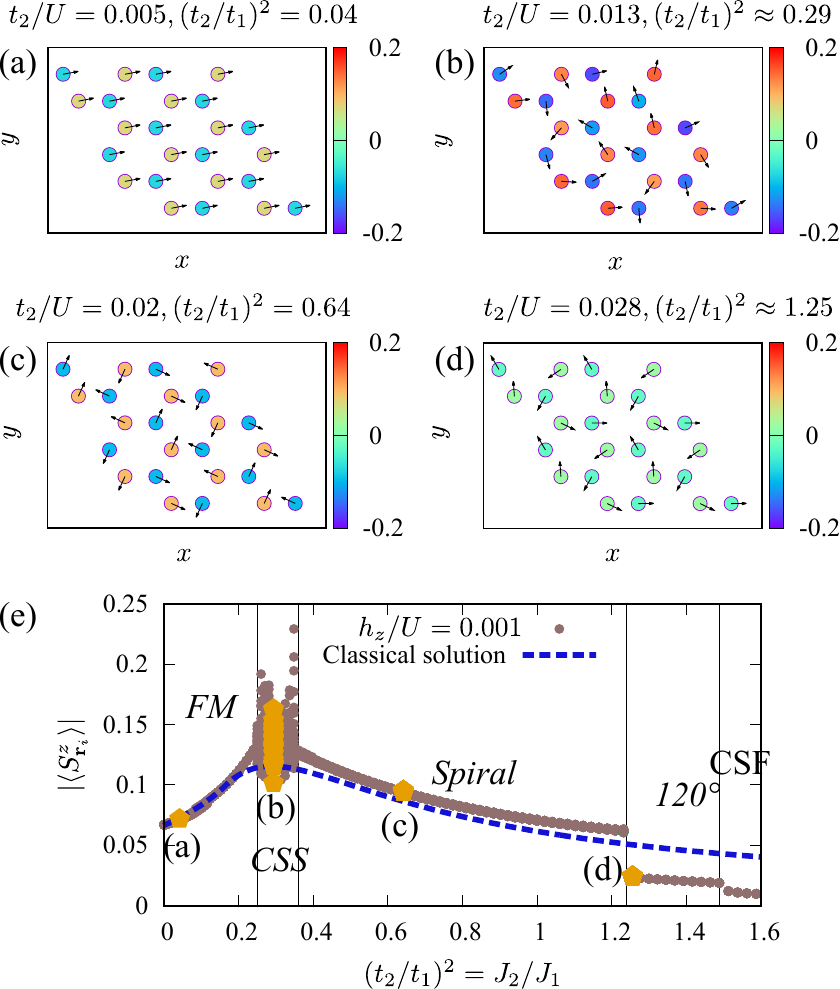}
  \caption{Results of the B-DMFT for different values of
    $\left( t_2/t_1 \right)^2 = J_2/J_1$ for $h_z/U = 10^{-3}$,
    $U_{\uparrow \downarrow}/U = 0.5, t_1/U = 0.025$ on a lattice of
    24 sites. \textbf{(a-d)} Different spin configurations. The color
    palette gives $\braket{S^z_{\vec{r}_i}}$, while arrows depict
    ordering in the $xy$-plane. \textbf{(a)} Uniform state with FM
    ordering; \textbf{(b)} CSS (chiral spin state) with no coplanar order; \textbf{(c)} A
    configuration of spiral states, in which each pseudo spin is
    aligned with only one of its three first neighbors and
    anti-aligned with two of its six second neighbors; \textbf{(d)} A
    $120^\circ$ configuration. \textbf{(e)} Absolute value of
    $\left|\braket{S^z_{\vec{r}_i}}\right|$. For each ratio
    $(t_2/t_1)^2$ we plot the result for all 24 sites and compare it
    to the classical solution. "Pentagons" mark results presented in
    (a-d). Note that for finite values of $h_z$ the border between the
    $120^\circ$ Mott state and CSF is slightly shifted in favour of
    the Mott state.}
  \label{fig:1}
\end{figure}

Deviations from this classical picture are already captured by B-DMFT
in the BKMH model.  In Fig.~\ref{fig:1}(a-d), we study the local
coplanar spin ordering (arrows), in the presence of an external
staggered magnetic field $h_z$, breaking the parity $\mc{P}$ symmetry
(reflection which maps the sublattice $A$ to the sublattice $B$):
\begin{equation}
  \label{eq:perturbationHz}
  H_z
  = h_z \Big( \sum\limits_{i\in{A}} S^z_{\vec{r}_i}
  -\sum\limits_{j\in{B}} S^z_{\vec{r}_j} \Big)
  \;.
\end{equation}
It corresponds to a staggered chemical potential in the boson language
and we will understand its role hereafter.  We directly infer some of
the ordered phases: at low $ J_2 / J_1 $, all spins are aligned in a
ferromagnetic (FM) order, while at large $ J_2 / J_1 $, we recover a
$120^\circ$ spiral order. For
$U_{\uparrow \downarrow}/U = 0.5, t_1/U = 0.025$ in the range
$0.36 \lesssim J_2/J_1 \lesssim 1.23$ we observe a different
configuration of spiral waves (Fig.~\ref{fig:1}(c)).  In addition, we
find an exotic intermediate regime when
$0.25 \lesssim J_2/J_1 \lesssim 0.36$ (we notice that positions of
phase boundaries are affected by $h_z$), characterized by a chiral
spin state (CSS) (this definition will be justified later) with no
coplanar magnetic order (Fig.~\ref{fig:1}(b)).  This is reminiscent of
the debated intermediate phase found in numerical studies on the XY
spin model~\cite{VarneySunGalitskiRigol2011,
  VarneySunGalitskiRigol2012, CarrasquillaEtAl2013, CioloEtAl2014,
  NakafujiIchinose2017, ZhuHuseWhite2013, ZhuWhite2014,
  BishopLiCampbell2014}.  On one hand, density matrix renormalization
group~\cite{ZhuHuseWhite2013, ZhuWhite2014} and coupled cluster
method~\cite{BishopLiCampbell2014} results evidenced an
antiferromagnetic Ising ordering along the $z$-axis, breaking $\mc{P}$
while preserving translational invariance. On the other hand, this
observation was not reported in ED~\cite{VarneySunGalitskiRigol2011,
  VarneySunGalitskiRigol2012} nor variational
Monte-Carlo~\cite{CarrasquillaEtAl2013, CioloEtAl2014,
  NakafujiIchinose2017} analyses, raising questions about the exact
nature of this intermediate phase.

Mapping the model onto a fermionic one and performing a mean-field
analysis~\cite{SedrakyanGlazmanKamenev2015, suppMaterial}, it was
proposed that an intermediate frustration stabilizes a phase with
spontaneously broken parity $\mc{P}$ and time-reversal $\mc{T}$
symmetries. This phase is characterized by antiferromagnetic
correlations and ChS fluxes staggered within the unit cell as in the
celebrated Haldane model~\cite{Haldane1988} and the authors suggested
that it realizes the chiral spin liquid state of
Kalmeyer-Laughlin~\cite{KalmeyerLaughlin1987, KalmeyerLaughlin1989}.
In this context, we plot in Fig.~\ref{fig:1}(e), the response for the
magnetization $\braket{S^z_{\vec{r}_i}}$ with respect to the field
$h_z$.  All phases except the CSS are characterized by a trivial
response to the perturbation: $\braket{S^z_{\vec{r}_i}} \sim h_z$,
whereas $\braket{S^z_{\vec{r}_i}}$ is strongly fluctuating in the CSS
(however we do not observe spontaneous symmetry breaking with
B-DMFT). These results cannot be explained in the context of a simple
coplanar ansatz, but could be related to a breaking of the degeneracy
between two mean-field solutions in the ChS field theory
description~\cite{suppMaterial}.

\begin{figure}
\includegraphics[width=0.4\textwidth]
{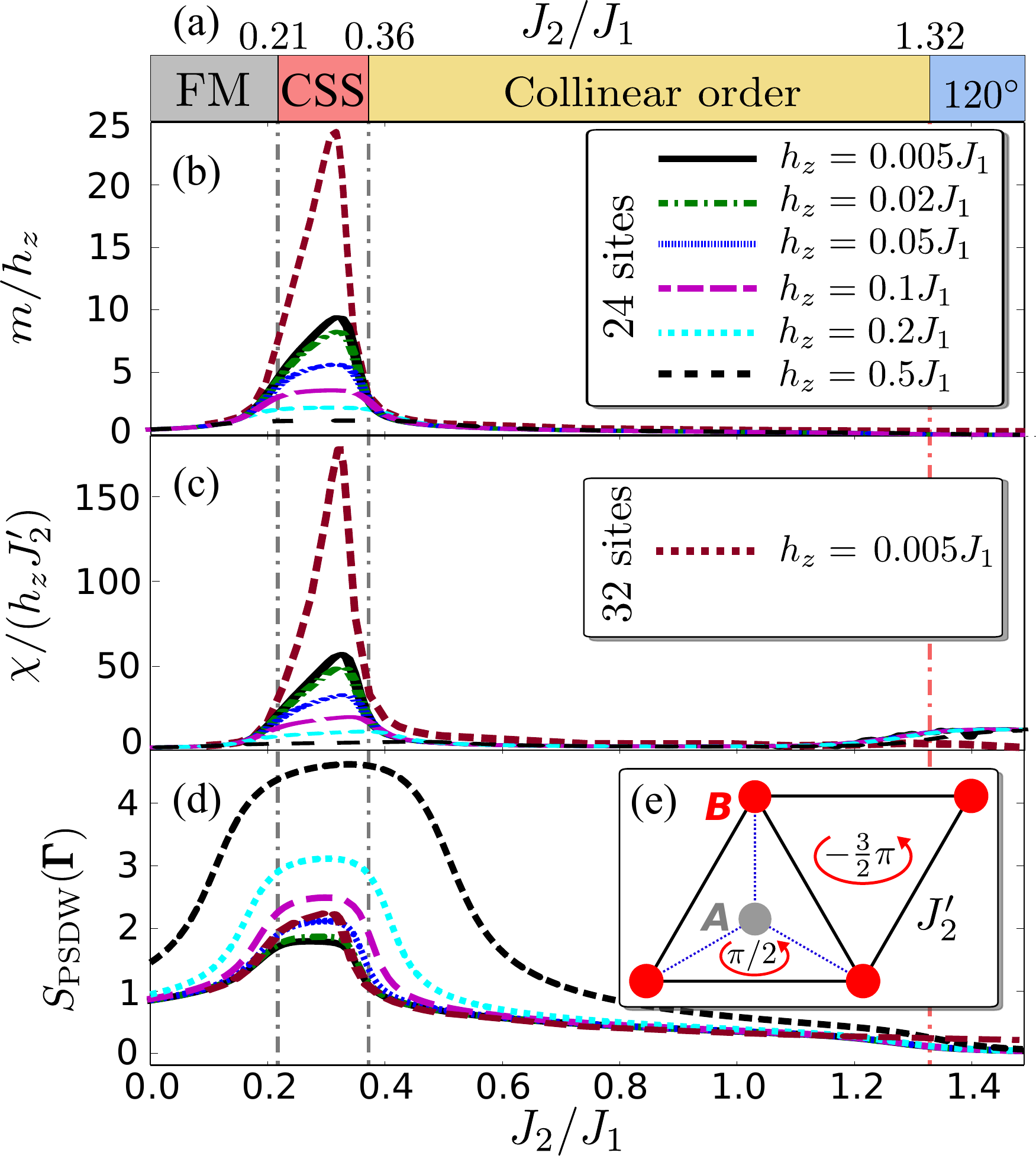}
\caption{\textbf{(a)} Phase diagram of the frustrated XY model from
  ED. \textbf{(b-d)} Variation of
  the observables with the dimensionless parameter $J_2/J_1$ for
  different values of $h_z$, with $J'_2 = 0.01 J_1$, on a lattice of
  $6 \times 2$ unit cells. \textbf{(b)} Difference of the average
  Ising magnetization on two sublattices $m$. \textbf{(c)} Scalar spin
  chirality $\chi$. \textbf{(d)} Pseudo-spin density wave structure factor
  $S_\text{PSDW}(\vec{\Gamma})$. \textbf{(e)} Schematic representation
  of the perturbation term $H_{J'_2}$.  }
\label{fig:2}
\end{figure}

\textit{II. ED on frustrated XY model.} We complete the study of the
effective frustrated XY model using ED and previously unaddressed
probes such as the responses to $\mc{P}$ and $\mc{T}$ breaking
perturbations and the topological description of the ground-state.  We
consider lattices of $24-32$ sites, with periodic boundary conditions,
and fixed total magnetization $S^z_\textrm{Tot} = 0$ if not stated
otherwise. First, we determine the phase boundaries using the fidelity
metric~\cite{ZanardiPaunkovic2006, ShiJian2010, VarneyEtAl2010,
  suppMaterial} $g$.  The phase diagram of the XY model deduced from
the ED calculations is given in Fig.~\ref{fig:2}(a). In agreement with
the B-DMFT analysis and previous numerical studies, we observe three
phase transitions at $J_2/J_1 \approx 0.21, 0.36$ and $1.32$. Small
deviations from the B-DMFT results could be due to a finite size of ED
clusters or non-perturbative interaction effects ($XY$ model does not
describe correctly the physics of the Mott phase when $t_i / U$ are
not small enough).  The nature of the phases detected with the ED is
verified by looking at the coplanar static structure factor
\begin{equation}
  \label{eq:sFactorSpiral}
  S_\textrm{Spiral}\left( \vec{q} \right) = 2 \sum\limits_{i, j \in A}
  e^{i\vec{q} \cdot \left( \vec{r}_i - \vec{r}_j \right)}
  \braket{S^x_{\vec{r}_i} S^x_{\vec{r}_j}} \;.
\end{equation}
Spiral waves display a maximum of $S_\textrm{Spiral}(\vec{q})$ at some
wave-vector(s) $\vec{q}$ in the first Brillouin zone. In the bosonic
language, this is interpreted as a macroscopic occupation of the
corresponding momentum state(s). We observe~\cite{suppMaterial} that
the phase in the region $J_2/J_1 \lesssim 0.21$ corresponds to the FM
order since $S_\textrm{Spiral}\left( \vec{q} \right)$ has a peak at
$\vec{q} = \vec{\Gamma}$. The phase at
$0.36 \lesssim J_2/J_1 \lesssim 1.32$ corresponds to a spiral wave
with collinear order (structure factor has maxima at three $\vec{M}$
points) as expected from the order by disorder mechanism. At
$1.32 \lesssim J_2/J_1$ the ground-state is the $120^\circ$ order
spiral wave (structure factor has a peak at two Dirac points
$\vec{K}$). In the intermediate frustration regime
($0.21 \lesssim J_2/J_1 \lesssim 0.36$) the coplanar static structure
factor is flat in the reciprocal space and we expect the ground-state
to be disordered in the $xy$-plane.  Notice that the ground-state in
all phases is located in the same sector of the total momentum at
point $\vec{\Gamma}$. Based on the ChS field theory predictions, the
order by disorder arguments and numerical observations, the CSS --
collinear order and collinear order -- $120^\circ$ order phase
transitions are expected to be of the first order, whereas the FM --
CSS phase transition -- of the second order.

As for the B-DMFT study, we analyze the linear response to external
perturbations breaking $\mc{P}$ and $\mc{T}$ symmetries. We are
interested in the relative magnetization between the two sublattices
$m = \Braket{m_{\vec{r}_i}} = \Braket{ S^z_{\vec{r}_i} -
  S^z_{\vec{r}_i + \vec{u}_3}}$, as well as the scalar spin chirality
$\chi = \Braket{\vec{S}_{\vec{r}_i} \cdot \left( \vec{S}_{\vec{r}_i +
      \vec{u}_1} \times \vec{S}_{\vec{r}_i + \vec{u}_2}
  \right)}$. Here we suppose that $i \in A$ and $\vec{u}_i$ are
vectors between first neighbor sites defined in Fig.~\ref{fig:0}(a).
When calculating the chirality $\chi$, we add a perturbation
corresponding to the second-neighbor hopping of the Haldane model, of
amplitude $J'_2$ and phase $\pi/2$ (as shown in Fig.~\ref{fig:2}(e)):
\begin{equation}
  \label{eq:perturbationDj2}
  H_{J'_2} = J'_2 \sum\limits_{\Braket{\Braket{ik}}} \big( e^{\pm i
    \pi/2} S_{\vec{r}_i}^+ S_{\vec{r}_k}^- + \hc \big)
  \;.
\end{equation}
We are interested in the limit $h_z, J'_2 \ll J_1$.  Results of the ED
calculations are presented in Figs.~\ref{fig:2}(b-c).  The CSS reveals
itself by sharp responses to such external fields.  Moreover, the
renormalized quantities $m / h_z$ and $\chi / (h_z J'_2)$ tend to
diverge in weak-coupling limit, giving a strong indication for
spontaneous symmetry breaking. This justifies our definition of the
CSS, which properties can be observed experimentally by tracking
on-site populations of bosons $n_{\sigma, \vec{r}_i}$ and currents
$J^\sigma_{ij} = \mf{Im}\Braket{b^\dag_{\sigma, \vec{r}_i} b_{\sigma,
    \vec{r}_j}}$\cite{Atala2014}.  One can probe the antiferromagnetic
ordering without breaking $\mc{P}$ and $\mc{T}$ by calculating the
pseudo-spin density wave (PSDW) structure
factor~\cite{VarneySunGalitskiRigol2011, VarneySunGalitskiRigol2012}:
\begin{equation*}
  S_\textrm{PSDW}(\vec{q}) =\!\! \sum\limits_{i,j} e^{i\vec{q} \cdot \left(
      \vec{r}_i - \vec{r}_j \right)}\Braket{m_{\vec{r_i}}  m_{\vec{r_j}}}.
\end{equation*}
We observe in Fig.~\ref{fig:2}(d) that $S_\textrm{PSDW}(\vec{q})$ has
a peak at $\vec{q} = \bm{\Gamma}$ in the intermediate frustration
regime.  These features are hardly affected by moderate Ising
interactions $K_i/J_1 \sim 0.1$ in
Eq.~\eqref{eq:xyModel}~\cite{Li2014}.

The observed spin configuration of the CSS could describe the chiral
spin liquid of Kalmeyer and Laughlin~\cite{KalmeyerLaughlin1987,
  KalmeyerLaughlin1989}.  Yet, we know that chiral spin liquids are
characterized by a topological degeneracy in the thermodynamic limit
on a compact space with genus $G$~\cite{WenWilczekZee1989, Wen1989,
  Wen1995}.  This property can be checked using ED in a system with
periodic boundaries: as $G = 1$ for a torus, one should have a
four-fold degenerate ground-state with two topological degeneracies
per chirality sector.  Still, because of finite size effects, one only
expects an approximate degeneracy in simulations.

\begin{figure}
  \includegraphics[width=0.48\textwidth] {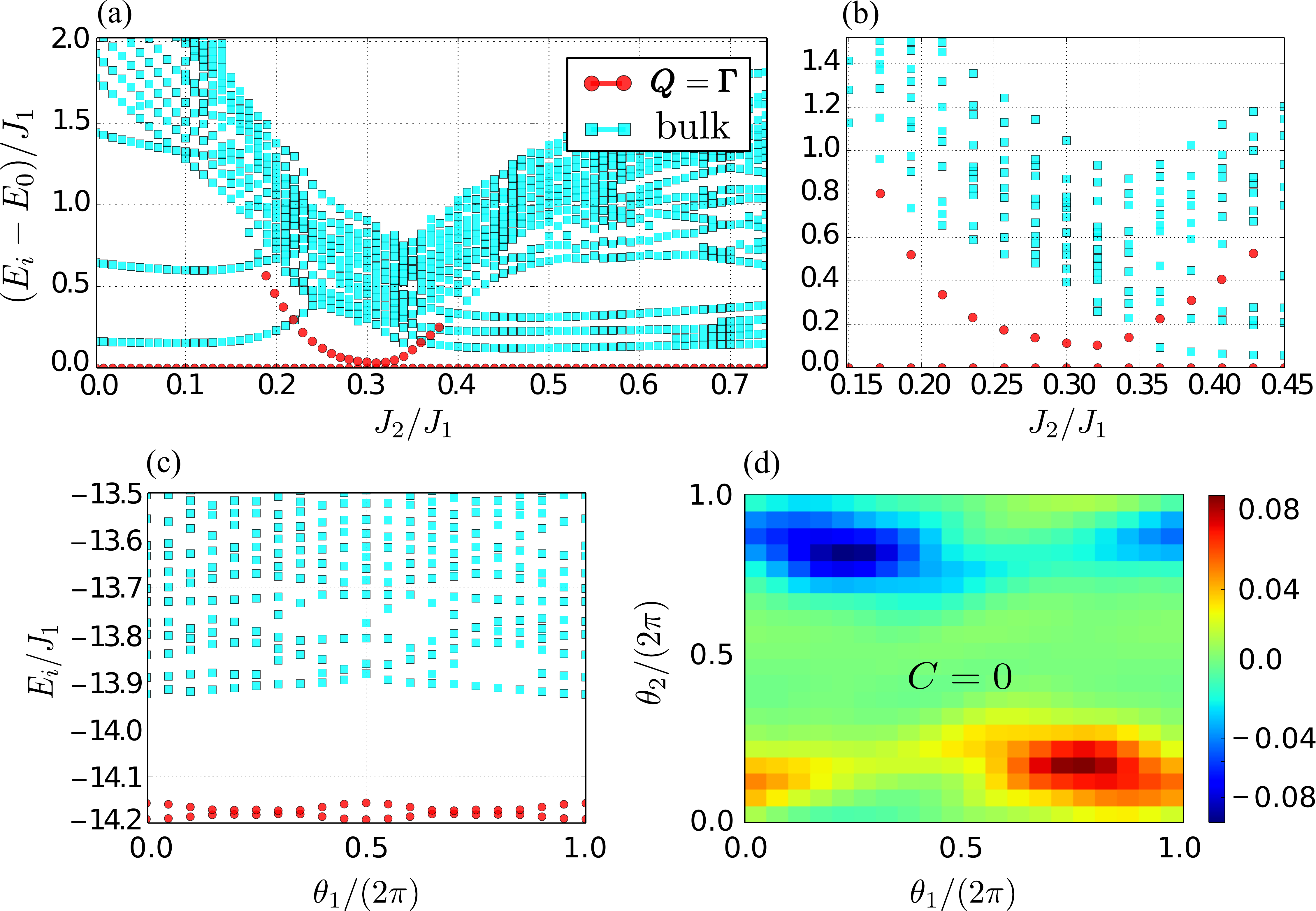}
  \caption{ED calculations of the low energy spectra as a function of
    $J_2/J_1$ \textbf{(a)} on a lattice of $4 \times 3$ unit cells for
    various $S^z_\text{Tot}$; \textbf{(b)} on a lattice of
    $4 \times 4$ unit cells in the $S^z_\text{Tot} = 0$ sector
    only. \textbf{(c)} Low energy spectrum as a function of the twist
    angle $\theta_1$ for $J_2/J_1 = 0.3$ and $\theta_2 = 0$ on a
    lattice of $4 \times 3$ unit cells. \textbf{(d)} Berry curvature
    calculated using the non-abelian formalism resulting in a
    vanishing Chern number shown for $J_2/J_1 = 0.3$,
    $h_z/J_1 = J_2'/J_1 = 0.02$ on a lattice of $4 \times 3$ unit
    cells.}
  \label{fig:3}
\end{figure}

In Fig.~\ref{fig:3}(a-b), we show the low-energy spectrum as a
function of $J_2/J_1$, resolved in different sectors of total momentum
$\vec{Q}$.  As mentioned previously, the ground-state always belongs
to the sector $\vec{Q} = \bm{\Gamma}$.  In the intermediate
frustration regime, we clearly observe the onset of a
doubly-degenerate ground-state manifold, well separated from higher
energy states.  The first excited state has the same momentum
$\vec{Q} = \bm{\Gamma}$, but lies in the opposite sector of
spin-inversion symmetry $S^z_{\vec{r}_i} \rightarrow -S^z_{\vec{r}_i}$
or reflection symmetry (that coincides with $\mc{P}$) for some
particular lattices. Low-lying excited state also moves away in energy when
the perturbations $H_z$ and $H_{J'_2}$ are switched on.

We probe the robustness of the low energy quasi-degenerate state
sector by performing the Laughlin's gedanken experiment and pumping a
quantum of magnetic flux through one of the non-trivial loops of the
torus~\cite{Laughlin1981, Laughlin1983, Thouless1989}. Numerically,
this is achieved using twisted boundary conditions in a translational
symmetry preserving manner. The results are given in
Fig.~\ref{fig:3}(c).  We observe that the same states in the sector
$\vec{Q} = \bm{\Gamma}$ are non-trivially gapped for all twists. For a
pumping of a single flux quantum we could not observe a crossing of
states in the ground-state manifold, that however does not imply that
the manifold is topologically trivial~\cite{WangGuCongSheng2011,
  HickeyCincioParamekanti2016, KumarChanglaniClarkFradkin2016}. The
topological nature of the ground-state manifold is unambiguously
determined by calculating the Chern number~\cite{NiuThoulessWu1985,
  Kohmoto1985, Hatsugai2004, Hatsugai2005}:
\begin{align}
  \label{eq:chernNumber}
  C = \frac{1}{2\pi} \int\limits_{0}^{2\pi} \int\limits_{0}^{2\pi}
  B(\theta_1, \theta_2) \textrm{d}\theta_1 \textrm{d}\theta_2 \;.
\end{align}
Here $\theta_1$ and $\theta_2$ are two angles of twisted boundary
conditions and $B(\theta_1, \theta_2)$ is the Berry
curvature~\cite{Berry1984}. We notice that two phases $\theta_i$
($i=1,2$) introduced in the spin language would correspond to four
phases $\theta^\sigma_i$ in the language of bosons of the BKMH model,
for which the spin component
$\theta^\uparrow_i - \theta^\downarrow_i = \theta_i$ is fixed and the
$U(1)$ component $\theta^\uparrow_i + \theta^\downarrow_i$ is
free~\cite{FuKane2006}.  Since the two quasi degenerate ground-states
lie in the same symmetry sector and cannot be separated unless twists
are trivial (reflection and spin-inversion symmetry
can not be used with twisted boundary conditions), we evaluate the
Berry curvature using the gauge-invariant non-abelian
formulation~\cite{YuQiBernevigFangDai2011, ShapourianClark2016,
  KrishnaHiteshClarkFradkin2016}:
$B(\theta_1, \theta_2) \delta \theta_1 \delta \theta_2 = \mf{Im} \ln
\mf{Det} \left( \mc{M}(\theta_1, \theta_2) \right)$, where elements of
the matrix $\mc{M}$ are obtained as follows:
\begin{align}
  \label{eq:berryCurvature}
  \mc{M}_{ij}(\theta_1, \theta_2)
  &=
    \Braket{\phi_i(\theta_1, \theta_2) | \phi_{\mu_1}(\theta_1 + \delta
    \theta_1, \theta_2)}
    \notag \\
  & \times
    \Braket{\phi_{\mu_1}(\theta_1 + \delta \theta_1, \theta_2) | \phi_{\mu_2}(\theta_1 + \delta
    \theta_1, \theta_2 + \delta \theta_2)}
    \notag \\
  & \times
    \Braket{\phi_{\mu_2}(\theta_1 + \delta \theta_1, \theta_2 + \delta
    \theta_2) | \phi_{\mu_3}(\theta_1, \theta_2 + \delta \theta_2)}
    \notag \\
  & \times
    \Braket{\phi_{\mu_3}(\theta_1, \theta_2 + \delta \theta_2) |
    \phi_j(\theta_1, \theta_2)}
    \;.
\end{align}
Here $\delta \theta_1$ and $\delta \theta_2$ refer to the numerical
mesh along the $\theta_1$ and $\theta_2$. $i, j, \mu_i = 1, 2$ are
indices of states $\Ket{\phi_1}$ and $\Ket{\phi_2}$ in the
ground-state manifold and the summation over $\mu_i$ is implicit. In
Fig.~\ref{fig:3}(d), we show a typical shape of the Berry
curvature. We find that the Chern number is zero in the intermediate
frustration regime. This result suggests that the intermediate phase
in the frustrated XY model is most likely to be a CSS with no
topological order, as suggested in Refs.~\onlinecite{ZhuHuseWhite2013,
  ZhuWhite2014, BishopLiCampbell2014} and not the Kalmeyer-Laughlin
state, with gauge fluctuations beyond the mean-field solution making
the phase topologically trivial as in the fermionic Kane-Mele model
case~\cite{RachelLehur2010, WuRachelLiuLehur2012, Hohenadler2012}.

To conclude, we studied the phase diagram of the bosonic Kane-Mele-Hubbard
model on the honeycomb lattice.  We have
shown that an effective frustrated XY model appears in the Mott
insulator phase. This model possesses an intermediate frustration
regime with a non-trivial chiral spin state, which breaks both
$\mc{P}$ and $\mc{T}$ symmetries.  It displays a finite scalar spin
chirality order and an antiferromagnetic ordering between
first-neighbor sites, while remaining translationally
invariant. Measuring the Chern number associated with this state
reveals its non-topological nature.

We thank Lo\"ic Herviou, Gr\'egoire Misguich, Stephan Rachel, C\'ecile
Repellin, Tigran Sedrakyan for insightful discussions. This work has
also benefitted from discussions at CIFAR meetings in Canada and
Soci\'et\'e Fran\c{c}aise de Physique.

Support by the Deutsche Forschungsgemeinschaft via DFG FOR 2414, DFG
SPP 1929 GiRyd, and the high-performance computing
center LOEWE-CSC is gratefully acknowledged.  This work was supported
in part by DAAD (German Academic and Exchange Service) under project
BKMH. I. V. acknowledges support by the Ministry of Education,
Science, and Technological Development of the Republic of Serbia under
projects ON171017 and BKMH, and by the European Commission under H2020
project VI-SEEM, Grant No. 675121. Numerical simulations were partly
run on the PARADOX supercomputing facility at the Scientific Computing
Laboratory of the Institute of Physics Belgrade. K. L. H.
acknowledges support from Labex PALM.



\bibliography{main.bbl}


\clearpage
\widetext
\begin{center}
  \textbf{\large Supplemental Material: Emergent Chiral Spin State in
    the Mott Phase of a Bosonic Kane-Mele-Hubbard Model}
\end{center}

\section{B-DMFT details}

For completeness, in this Section we briefly describe the B-DMFT
method along the lines of references~\cite{HubenerSnoekHofstetter2009,
  Snoek&Hofstetter, AndersGullPolletTroyerWerner2010, LiHof}. In particular, in order to be able to address exotic states that break translational invariance, we implement real-space B-DMFT \cite{Snoek&Hofstetter,LiHof,HeLiHof, HeJiHof}. The
essence of DMFT is mapping of the full lattice model onto a set of
local models whose parameters are determined through a
self-consistency condition. The self-consistency is imposed on the
level of single--particle Green's functions that can be written in the
Nambu notation as
\begin{equation}
 G_{ij}(\tau, \eta)\equiv  G_{ij}(\tau-\eta)= -T_{\tau, \eta}\left\langle\left(\begin{array}{cccc}
 b_{\uparrow,\vec{r}_i}(\tau) b_{\uparrow,\vec{r}_j}^{\dagger} (\eta) & b_{\uparrow,\vec{r}_i}(\tau) b_{\uparrow,\vec{r}_j}(\eta) & b_{\uparrow,\vec{r}_i}(\tau) b_{\downarrow,\vec{r}_j}^{\dagger}(\eta) & b_{\uparrow, \vec{r}_i}(\tau) b_{\downarrow,\vec{r}_j}(\eta)\\
  b_{\uparrow,\vec{r}_i}^{\dagger}(\tau) b_{\uparrow,\vec{r}_j}^{\dagger} (\eta) & b_{\uparrow,\vec{r}_i}^{\dagger}(\tau) b_{\uparrow,\vec{r}_j}(\eta) & b_{\uparrow,\vec{r}_i}^{\dagger}(\tau) b_{\downarrow, \vec{r}_j}^{\dagger}(\eta) & b_{\uparrow, \vec{r}_i}^{\dagger}(\tau) b_{\downarrow, \vec{r}_j}(\eta)\\
   b_{\downarrow,\vec{r}_i}(\tau) b_{\uparrow, \vec{r}_j}^{\dagger}(\eta)& b_{\downarrow, \vec{r}_i}(\tau) b_{\uparrow, \vec{r}_j}(\eta) & b_{\downarrow, \vec{r}_i}(\tau)b_{\downarrow, \vec{r}_j}^{\dagger}(\eta) & b_{\downarrow, \vec{r}_i}(\tau) b_{\downarrow, \vec{r}_j}(\eta)\\
      b_{\downarrow,\vec{r}_i}^{\dagger}(\tau) b_{\uparrow, \vec{r}_j}^{\dagger}(\eta)& b_{\downarrow, \vec{r}_i}^{\dagger}(\tau) b_{\uparrow, \vec{r}_j}(\eta) & b_{\downarrow, \vec{r}_i}^{\dagger}(\tau)b_{\downarrow, \vec{r}_j}^{\dagger}(\eta) & b_{\downarrow, \vec{r}_i}^{\dagger}(\tau) b_{\downarrow, \vec{r}_j}(\eta)
 \end{array}\right)\right\rangle \;.
\end{equation}
In the following we express the Green's functions in terms of Matsubara
frequencies $\omega_n=2\pi n/\beta$, where $\beta $ is the inverse
temperature (in the zero temperature limit $\beta\rightarrow\infty$) and
$G_{ij}(i\omega_n) = \int\, d{\tau} \exp(i \omega_n \tau)
G_{ij}(\tau)$. 

In real-space B-DMFT we decompose the full lattice problem into a set
of local single-site effective problems. The approximation is such
that local correlations are fully taken into account, while non-local
correlations are treated at the mean-field level. At each site $i$, we
attach a bath described by orbital degrees of freedom.  The effective
local Hamiltonian is given by a bosonic Anderson impurity (AI) model
\cite{Snoek&Hofstetter}
\begin{eqnarray}
  \mathcal{H}^\mathrm {AI}_i
  &=& \sum_{l=0}^L \left[\varepsilon_l
      a_l^{\dagger}  a_l +
      \sum_{\sigma}\left(V_{l, \sigma}
      a_l^{\dagger}  b_{\sigma, \vec{r}_i} +
      V_{l, \sigma}^*  a_l  b_{\sigma,
      \vec{r}_i}^{\dagger} +  W_{l, \sigma}  a_l
      b_{\sigma, \vec{r}_i} + W_{l, \sigma}^*
      a_l^{\dagger}  b_{\sigma,
      \vec{r}_i}^{\dagger}\right)\right]
      \nonumber\\
  &+& \sum_{\sigma}\left(-\psi_{\sigma, \vec{r}_i}^{\mathrm {AI}*}  b_{\sigma,
      \vec{r}_i} - \psi_{\sigma, \vec{r}_i}^{\mathrm {AI}}  b^{\dagger}_{\sigma, \vec{r}_i} +
      \frac{U}{2}  n_{\sigma, \vec{r}_i} ( n_{\sigma, \vec{r}_i}-1)-\mu_{\sigma}
      n_{\sigma, \vec{r}_i}\right) + U_{\uparrow \downarrow} n_{\uparrow, \vec{r}_i} n_{\downarrow,
      \vec{r}_i},
\label{eq:Anderson}
\end{eqnarray}
where the index $l$ labels the Anderson orbitals with energies
$\varepsilon_l$ and we allow for complex values of the Anderson
parameters $V_{l, \sigma}$ and $W_{l, \sigma}$ that couple orbital
degrees of freedom with impurity atoms. We use $L=4$; we check that
results are the same for $L=5$ and $6$. Local interaction terms
proportional to $U$ and $U_{\uparrow \downarrow}$ come directly from
the initial lattice model and, as we work in the grand canonical
ensemble, we introduce chemical potentials $\mu_{\sigma}\equiv\mu$. We
define hybridization functions of the Anderson impurity model as
\begin{eqnarray}
  \Delta_{11}^{\nu \mu}(i \omega_n)
  &=& \sum_l\left(\frac{V_{l, \nu
      }^*V_{l, \mu
      }}{\epsilon_l-i\omega_n}+\frac{W_{l, \nu
      }^*W_{l, \mu
      }}{\epsilon_l+i\omega_n}\right),\\
  \Delta_{22}^{\nu \mu}(i \omega_n)
  &=& \sum_l\left(\frac{W_{l, \mu
      }^*W_{l, \nu
      }}{\epsilon_l-i\omega_n}+\frac{V_{l, \mu
      }^*V_{l, \nu
      }}{\epsilon_l+i\omega_n}\right),\\
  \Delta_{12}^{\nu \mu}(i \omega_n)
  &=& \sum_l\left(\frac{V_{l, \nu
      }^*W^*_{l,\mu
      }}{\epsilon_l-i\omega_n}+\frac{V_{l, \mu
      }^*W^*_{l, \nu
      }}{\epsilon_l+i\omega_n}\right),\\
  \Delta_{21}^{\nu \mu}(i \omega_n)
  &=& \sum_l\left(\frac{V_{l, \mu}W_{l, \nu}
      }{\epsilon_l-i\omega_n}+\frac{V_{l,\nu
      } W_{l, \mu
      }}{\epsilon_l+i\omega_n}\right), 
 \label{eq:hyb}
\end{eqnarray}
and introduce a $4 \times 4 $ matrix $\Delta (i \omega_n) $ as
\begin{equation}
  \Delta (i \omega_n) \equiv \left(
    \begin{array}{cccc}
      \Delta_{11}^{\uparrow \uparrow} &  \Delta_{12}^{\uparrow \uparrow} & \Delta_{11}^{\uparrow \downarrow}& \Delta_{12}^{\uparrow \downarrow}\\
      \Delta_{21}^{\uparrow \uparrow} &  \Delta_{22}^{\uparrow \uparrow} & \Delta_{21}^{\uparrow \downarrow}& \Delta_{22}^{\uparrow \downarrow}\\
      \Delta_{11}^{\downarrow \uparrow}& \Delta_{12}^{\downarrow \uparrow} & \Delta_{11}^{\downarrow \downarrow} &  \Delta_{12}^{\downarrow \downarrow}\\
      \Delta_{21}^{\downarrow \uparrow}& \Delta_{22}^{\downarrow \uparrow} & \Delta_{21}^{\downarrow \downarrow} &  \Delta_{22}^{\downarrow \downarrow}
    \end{array}
  \right).
\end{equation}
The following relations for the hybridization functions hold true:
\begin{eqnarray}
  \Delta_{22}^{\uparrow\uparrow}(i\omega_n) =
  		\Delta_{11}^{{\uparrow\uparrow}*}(i\omega_n), \quad
  \Delta_{21}^{\uparrow\uparrow}(i\omega_n) = 
  		\Delta_{12}^{{\uparrow\uparrow}*}(i\omega_n), \quad
  \Delta_{22}^{\downarrow\downarrow}(i\omega_n) =
  		\Delta_{11}^{{\downarrow\downarrow}*}(i\omega_n), \quad
  \Delta_{21}^{\downarrow\downarrow}(i\omega_n) = 
  		\Delta_{12}^{{\downarrow\downarrow}*}(i\omega_n), \nonumber\\
  \Delta_{22}^{\uparrow\downarrow}(i\omega_n) = 
  		\Delta_{11}^{{\uparrow\downarrow}*}(i\omega_n), \quad
  \Delta_{21}^{\uparrow\downarrow}(i\omega_n) = 
  		\Delta_{12}^{{\uparrow\downarrow}*}(i\omega_n), \quad
  \Delta_{22}^{\downarrow\uparrow}(i\omega_n) = 
  		\Delta_{11}^{{\downarrow\uparrow}*}(i\omega_n), \quad
  \Delta_{21}^{\downarrow\uparrow}(i\omega_n) = 
  		\Delta_{12}^{{\downarrow\uparrow}*}(i\omega_n).
\end{eqnarray}
The terms $\psi_{\sigma, \vec{r}_i}^{\mathrm{AI}}$ used in
Eq.~(\ref{eq:Anderson}) incorporate a correction with respect to the
mean--field result and they read~\cite{Snoek&Hofstetter}:
\begin{eqnarray}
  \psi_{\uparrow,\vec{r}_i}^{\mathrm{ AI}}
  &=& \sum_j t_{\uparrow, ij}\phi_{\uparrow, \vec{r}_j}-\phi_{\uparrow, \vec{r}_i}^*\Delta_{21}^{\uparrow\uparrow}(0)-\phi_{\downarrow, \vec{r}_i}^*\Delta_{21}^{\downarrow\uparrow}(0)-\phi_{\uparrow, \vec{r}_i} \Delta_{11}^{\uparrow\uparrow}(0)-\phi_{\downarrow, \vec{r}_i} \Delta_{11}^{\downarrow\uparrow}(0),\\
  \psi_{\downarrow,\vec{r}_i}^{\mathrm{ AI}}
  &=& \sum_j t_{\downarrow, ij}\phi_{\downarrow, \vec{r}_j}-\phi_{\uparrow,\vec{r}_i}^*\Delta_{21}^{\uparrow\downarrow}(0)-\phi_{\downarrow,\vec{r}_i}^*\Delta_{21}^{\downarrow\downarrow}(0)-\phi_{\uparrow, \vec{r}_i}
      \Delta_{11}^{\uparrow\downarrow}(0)-\phi_{\downarrow, \vec{r}_i} \Delta_{11}^{\downarrow\downarrow}(0),
\end{eqnarray}
where the condensate order parameters are defined as
\begin{equation}
  \phi_{\sigma, \vec{r}_i} = \langle b_{\sigma, \vec{r}_i}\rangle,
  \label{eq:con_order}
\end{equation} 
and $t_{\sigma,ij} $ are hopping amplitudes of the two species defined
in the initial lattice model.

By exact diagonalization of the local model (\ref{eq:Anderson}) we
obtain the values of the local Green's functions. From here, the local
self--energy is obtained from the local Dyson equation
\begin{equation}
  (G^{\mathrm{AI}})^{-1}_{ii} (i \omega_n) =  
  \left(
    \begin{array}{cccc}
      i \omega_n +\mu& & &\\
                     & -i \omega_n + \mu & &\\
                     & &i \omega_n +\mu & \\
                     & & & -i \omega_n + \mu
    \end{array}
  \right)+ \Delta(i\omega_n) - \Sigma_i^{\mathrm{AI}}.
 \label{eq:localdysonequation}
\end{equation}
The approximate real-space Dyson equation takes the following form:
\begin{equation}
  G^{-1}_{ij, \mathrm{latt}} (i \omega_n) =  \left(
    \begin{array}{cccc}
      \!\!\left(i \omega_n +\mu  \right)\delta_{ij}+ t_{\uparrow,ij}& & &\\
      \! & \!\!\left(-i \omega_n + \mu \right)\delta_{ij}+t^*_{\uparrow, ij}\!\! &\!&\!\\
      \! &\!&\!\!\left(i \omega_n \!+\!\mu \right)\delta_{ij}+ t_{\downarrow, ij}& \!\!\\
      \! &\!&\!& \left(-i \omega_n \!+\! \mu \right)\delta_{ij}+t^*_{\downarrow, ij}\\
    \end{array}\right)-\delta_{ij} \Sigma_i^{\mathrm{AI}},
  \label{eq:realspacedyson}
\end{equation}
where we approximate the self--energy by a local contribution from
Eq.~(\ref{eq:localdysonequation}). The last step represents the main
approximation of DMFT.  Finally, we need a criterion to set the values of
the parameters $\varepsilon_l$, $V_{l, \sigma}$ and $W_{l, \sigma}$ in
Eq.~(\ref{eq:Anderson}).  To this end, a condition is imposed on the
hybridization functions (\ref{eq:hyb}).  These functions should be
optimized such that the two Dyson equations,
(\ref{eq:localdysonequation}) and (\ref{eq:realspacedyson}), yield the
same values of the local Green's functions. In practice, we iterate a
self--consistency loop to fulfill this condition, starting from
arbitrary initial values. At the same time we impose a simple self
consistency on the local condensate order parameters
$\phi_{\sigma, \vec{r}_i}$.

Once that the self-consistency is achieved and values of Anderson
parameters $\varepsilon_l, V_{l, \sigma}$ and $W_{l, \sigma}$ are
fixed, by solving the local model (\ref{eq:Anderson}) we obtain
results for local condensate order parameters (\ref{eq:con_order}) and
the expectation values of the pseudo spin operators
\begin{eqnarray}
  \langle S^x_{{\mathbf r}_i} \rangle
  &=& \langle b_{\uparrow,\vec{r}_i}^{\dagger} b_{\downarrow,\vec{r}_i}+b_{\downarrow,\vec{r}_i}^{\dagger} b_{\uparrow,\vec{r}_i}\rangle/2,
      \label{eq:spin_order_x}\\
  \langle S^y_{{\mathbf r}_i} \rangle
  &=& i \langle b_{\uparrow,\vec{r}_i}^{\dagger} b_{\downarrow,\vec{r}_i}-b_{\downarrow,\vec{r}_i}^{\dagger} b_{\uparrow,\vec{r}_i}\rangle/2,
      \label{eq:spin_order_y}\\
  \langle S^z_{{\mathbf r}_i} \rangle
  &=& \langle b_{\uparrow,\vec{r}_i}^{\dagger} b_{\uparrow,\vec{r}_i}-b_{\downarrow,\vec{r}_i}^{\dagger} b_{\downarrow,\vec{r}_i}\rangle/2.
      \label{eq:spin_order_z}
\end{eqnarray}
We work with a finite lattice consisting of 96 sites and periodic
boundary conditions that provide a proper sampling of the Brillouin
zone that includes its corners~\cite{VarneyEtAl2010}. The values of the chemical potential terms in Eq.~(\ref{eq:Anderson}) are fixed to
$\mu_{\sigma} = U_{\uparrow\downarrow}/2$.

\begin{figure}
  \includegraphics[width=\linewidth]{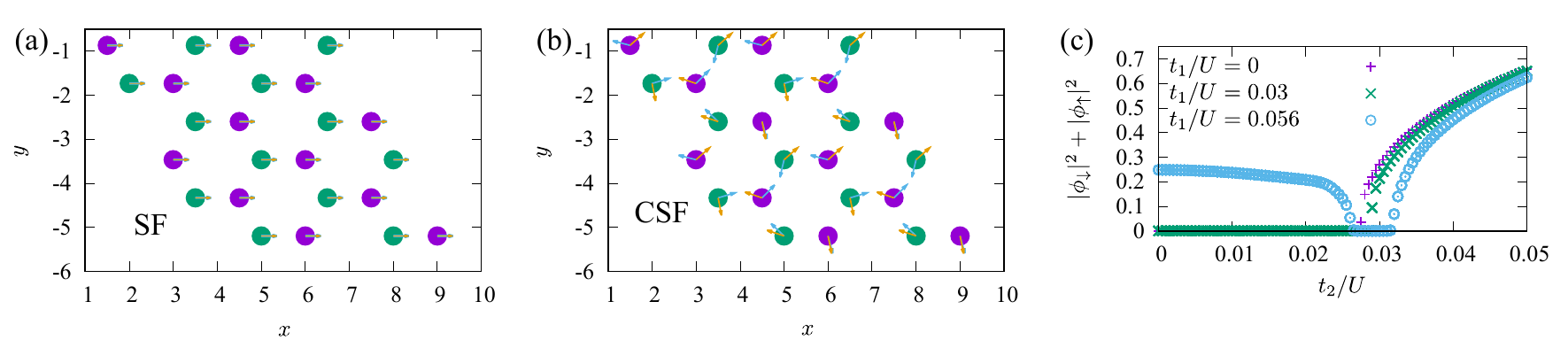}
  \caption{Color maps: Real-space distribution of condensate order
    parameters of the two bosonic species in (a) uniform superfluid
    (SF)
    ($t_1/U = 0.056, t_2/U = 0.005, U_{\uparrow\downarrow}/U = 0.5,
    \mu/U_{\uparrow \downarrow} = 0.5$), and (b) chiral superfluid
    (CSF)
    ($t_1/U = 0.018, t_2/U = 0.032, U_{\uparrow\downarrow}/U = 0.5,
    \mu/U_{\uparrow \downarrow} = 0.5$). Local condensate order
    parameters are aligned in the SF. In contrast, they exhibit
    $2\pi/3$ winding in the CSF. The "winding direction" is opposite
    for the two species and for the two sublattices, implying that for
    each sublattice the two species condense in the two different
    Dirac points. The choice of the Dirac points is opposite for the
    two sublattices. (c) The condensate order parameters as a function of $t_2/U$ for several values of $t_1/U$.}
  \label{Fig:Figsf}
\end{figure}

Finite values of condensate order parameters (\ref{eq:con_order}) mark
a superfluid phase, while vanishing values correspond to a Mott
insulator state (MI).  We further distinguish a uniform superfluid
(SF), where the order parameters of the two species on both
sublattices are aligned, Fig.~\ref{Fig:Figsf}(a), and a chiral
superfluid (CSF) with $2\pi/3$ winding of the order parameters,
Fig.~\ref{Fig:Figsf}(b).  For the parameters studied in the paper, we
find that the absolute values of the order parameters are the same for
the two species on all lattice sites, yet for CSF state winding
directions are opposite for the two species on the two sublattices,
Fig.~\ref{Fig:Figsf}(b). Moreover, in CSF phase condensate order
parameters on the two sublattices and for the two species are
determined up to a relative phase, Fig.~\ref{Fig:Figsf}(b).  We also
expect that similarly to the case of the bosonic Haldane
model~\cite{VasicEtAl2015} the SF -- CSF phase transition is of the
first order, whereas the SF (CSF) -- MI phase transition is of the
second order. 

In Fig.~\ref{Fig:Figsf}(c) we  plot absolute values of the order parameters (\ref{eq:con_order}) (which are uniform throughout the lattice) as functions of $t_2/U$ for several values of $t_1/U$. For the case of $t_1/U = 0$, we find a transition from the Mott state into the chiral superfluid state at $ t_2/U\approx 0.027$. At $t_1/U = 0.03$, the transition sets in at a slightly higher value $t_2/U \approx 0.0285$. The most interesting behavior is found for  $t_1/U = 0.056$, where for small values of $t_2$ we find a uniform superfluid. With an increase in $t_2$, at $t_2/U \approx 0.0265 $ the 
Mott insulator state is reached due to competing effects of $t_1$ and $t_2$, and finally at $t_2/U \approx 0.0315$ the system becomes a chiral superfluid. These results are summarized in the phase diagram of BKMH model (Fig.~1(b) of the main text).

Different magnetic orderings within the Mott domain, as
discussed in Fig.~1, are distinguished based on the order parameters
defined in Eqs.~(\ref{eq:spin_order_x}) and
(\ref{eq:spin_order_y}). In Fig.~2 we show the results of a
calculation on a 24-site lattice. We monitor magnetic ordering in
$z$-direction marked by finite values of order parameter
(\ref{eq:spin_order_z}) that are introduced by a finite value of $h_z$
as defined in equation (\ref{eq:perturbationHz}). We have checked that a four-fold increase in lattice size (96 vs.~24 lattice sites) introduces a shift in the position of the "intermediate region" borders of the order of $\Delta t_2/U \sim 2 \times 10^{-4} $ or less than $2\%$ in 
relative units.

\section{Classical solution}

We consider an ansatz for the classical ($S \rightarrow \infty$)
solution of the spin problem defined as follows:
\begin{equation}
  \label{eq:classicalAnsatz}
  \bm{S}_{\vec{r}_i} = S
  \begin{pmatrix}
    \sin \left( \theta_\mu \right) \cos \left( \phi_{\mu, i} \right) \\
    \sin \left( \theta_\mu \right) \sin \left( \phi_{\mu, i} \right) \\
    \cos \left( \theta_{\mu} \right)
  \end{pmatrix}
  \;.
\end{equation}
Here $\mu \in [A, B]$ is the sublattice index and a free parameter
$\theta_{\mu}$ characterizes the orientation of the spin on the
sublattice $\mu$ with respect to the $z$-axis. It verifies
$0 \leq \theta_{\mu} \leq \pi$ ($\sin \theta_\mu$ is always
positive). Similarly to the
Refs.~\onlinecite{RastelliTassiReatto1979,MudlerEtAl2010}, we define
phases $\phi_{A, i} = \bm{q}\cdot\bm{R}_i$ and
$\phi_{B, i} = \bm{q}\cdot\bm{R}_i + \eta$, where $\vec{q}$ is the
spiral wave vector and $\eta$ describes the relative orientation of
spins on sublattices $A$ and $B$ at the same unit cell. The (anti-)
ferromagnetic ordering between first-neighbor sites in the $XY$-plane
is thus described by $\bm{q} = 0$, $\eta = 0(\pi)$ and
$\theta_\mu = \pi/2$. The Ising antiferromagnetic ordering is defined
by $\theta_A = 0$, $\theta_B = \pi$ and its $\mb{Z}_2$ symmetric
solution $\theta_A = \pi$, $\theta_B = 0$.

\subsection{Zero external magnetic field $h_z$}

We write the energy per spin in terms of parameters of the Hamiltonian
$H$ in Eq.~\eqref{eq:xyModel} for $K_i = 0$:
\begin{align}
  \epsilon =
  &
    - J_1 S^2 \sin \theta_A \sin \theta_B \left[
    \cos \eta + \cos \left( \eta - Q_1 \right) + \cos \left( \eta + Q_2 \right) \right]
    \notag \\
  &
    + J_2 S^2 \left( \sin^2 \theta_A + \sin^2 \theta_B \right) \left[
    \cos Q_1 + \cos Q_2 + \cos \left( Q_1 + Q_2 \right) \right] \;.
\end{align}
Here for simplicity we defined $Q_i = \vec{q} \cdot \bm{v}_i$ with
$\vec{v}_i$ -- 3 second-neighbor vectors. By minimizing the energy per
spin with respect to all the parameters that we introduced, we obtain
that only coplanar solutions with $\theta_\mu = \pi/2$ will
survive. In this case we
recover~\cite{RastelliTassiReatto1979,MudlerEtAl2010}
\begin{align}
  \label{eq:spiralWaveSolution}
  & \cos \eta = \frac{2 J_2}{J_1} \left( 1 + \cos Q_1 + \cos Q_2 \right)
    \;, \notag \\
  & \sin \eta = \frac{2 J_2}{J_1} \left( \sin Q_1 - \sin Q_2 \right)
    \;, \notag \\
  & \cos Q_1 + \cos Q_2 + \cos \left( Q_1 + Q_2 \right) =
    \frac{1}{2} \left( \frac{J_1^2}{4J_2^2} - 3 \right)
    \;.
\end{align}
The uniform solution at $\bm{q} = \vec{\Gamma}$ and $\eta = 0$ is
valid until $J_2 / J_1 \leq 1/6$. Spiral waves solution is valid in
the regime $J_2 / J_1 > 1/6$ for $J_1 \neq 0$. When two sublattices
are decoupled ($J_1 = 0$), the solution corresponds to the $120^\circ$
order.  The energy per spin of the uniform solution is
$\epsilon_\text{cl} = - 3 S^2 \left( J_1 - 2 J_2 \right)$, whereas the
energy corresponding to the spiral wave state is
$\epsilon_\text{sp} = - S^2 J_1 \left( \frac{J_1}{4J_2} +
  \frac{3J_2}{J_1} \right)$.

\subsection{Effect of the external magnetic field $h_z$}

Now we are interested in the effect of the external magnetic field
$h_z$ on the stabilization of the out-of plane (PSDW) solution. We
calculate the energy per spin when the perturbation term $H_z$ of
Eq.~\eqref{eq:perturbationHz} is added to the Hamiltonian:
\begin{align}
  \epsilon =
  & - J_1 S^2 \sin \theta_A \sin \theta_B \left[
    \cos \eta + \cos \left( \eta - Q_1 \right) + \cos
    \left( \eta + Q_2 \right) \right]
    \notag \\
  & + J_2 S^2 \left( \sin^2 \theta_A + \sin^2
    \theta_B \right) \left[
    \cos Q_1 + \cos Q_2 + \cos \left( Q_1 + Q_2 \right) \right]
    - \frac{h_z S}{2} \left( \cos \theta_A - \cos \theta_B  \right)
    \;.
\end{align}
We suppose that the angle $\theta_\mu$ is close to $\pi/2$ (the
solution is almost coplanar) for small values of $h_z$ and we perform
the expansion in powers of $\tilde{\theta}_\mu = \pi/2 -
\theta_\mu$. At the first order in the expansion we observe that the
coplanar degree of freedom and the degree of freedom along the
$z$-axis become decoupled. Values of $\eta$, $Q_1$ and $Q_2$
correspond to the spiral wave solution~\eqref{eq:spiralWaveSolution}
and parameters $\tilde{\theta}_A$ and $\tilde{\theta}_B$ are deduced
using the following relation:
\begin{align}
  & \tilde{\theta}_A + \tilde{\theta}_B = 0
    \;, \notag \\
  & \tilde{\theta}_A - \tilde{\theta}_B = \frac{h_z}{J_1 \left[ \cos \eta + \cos
    \left( \eta - Q_1 \right) + \cos \left( \eta + Q_2 \right) \right]
    - 2 J_2 \left[ \cos Q_1 + \cos Q_2 +
    \cos \left( Q_1 + Q_2 \right) \right]}
    \;.
\end{align}
In the regime $J_2 / J_1 \leq 1/6$ we obtain
\begin{equation}
  \tilde{\theta}_A = - \tilde{\theta}_B = \frac{h_z}{6 \left( J_1 - 2 J_2 \right)}
  \;,
\end{equation}
and in the regime $J_2 / J_1 > 1/6$
\begin{equation}
  \tilde{\theta}_A = - \tilde{\theta}_B = \frac{2 h_z J_2}{
    \left( J_1^2 + 12 J_2^2 \right)}
  \;.
\end{equation}
We see thus that for the classical ansatz~\eqref{eq:classicalAnsatz}
the linear response of the spin to the applied magnetic field $h_z$ is
supposed to be small and of the order of $h_z$.

\section{Mean-field solution and the ChS field theory description}

According to the Ref.~\onlinecite{SedrakyanGlazmanKamenev2015} one can
preform a mapping of the spin problem~\eqref{eq:xyModel} onto the
problem of spinless fermions coupled to ChS gauge
fields~\cite{Fradkin1989, AnbjornSemenoff1989, LopezRojoFradkin1994,
  Misguichjolicoeurgirvin2001, SunKumarFradkin2015}. At the mean-field
level, the system is stabilized in the chiral spin state by forming
the anti-ferromagnetic order and staggered ChS fluxes within the unit
cell identical to the fluxes of the Haldane
model~\cite{Haldane1988}. This allowed authors of the
Ref.~\onlinecite{SedrakyanGlazmanKamenev2015} to suggest that the
resulting solution (that breaks spontaneously $\mc{P}$ and $\mc{T}$ symmetries)
could be a chiral spin liquid state of Kalmeyer-Laughlin and deduce
the phase boundaries, that were in good agreement with the numerical
data~\cite{VarneySunGalitskiRigol2011, VarneySunGalitskiRigol2012,
  CarrasquillaEtAl2013, CioloEtAl2014, NakafujiIchinose2017,
  ZhuHuseWhite2013, ZhuWhite2014, BishopLiCampbell2014}. Below, we
represent analytical arguments that lead to this suggestion.

\subsection{Zero external magnetic field $h_z$}

The problem of the Eq.~\eqref{eq:xyModel} can be rewritten in the
fermionic language using the following transformation:
\begin{equation}
  S^+_{\vec{r}_j} = c^\dag_{\vec{r}_j} e^{i\alpha_{\vec{r}_j}}, \quad
  \alpha_{\vec{r}_j} = \sum\limits_{k \neq j} B_{jk} n_{\vec{r}_k}, \quad
  n_{\vec{r}_k} = c^\dag_{\vec{r}_k} c_{\vec{r}_k} = S^z_{\vec{r}_k} + 1/2  \;.
\end{equation}
Here $c^\dag_{\vec{r}_j}$ and $c_{\vec{r}_j}$ are fermionic creation
and annihilation operators and
\begin{equation}
  B_{jk} = \text{arg}\left( \tau_k - \tau_j \right) =
  \mf{Im} \ln \left( \tau_k - \tau_j \right) \;,
\end{equation}
with the complex number $\tau_j = x_{j} + i y_{j}$ associated to each
point on the lattice defined by the vector
$\bm{r}_j = x_{j}\bm{e}_x + y_{j}\bm{e}_y$. $B_{jk}$ could be
interpreted as the angle that the vector $\bm{r}_k - \bm{r}_j$ forms
with the $x$-axis. The Hamiltonian~\eqref{eq:xyModel} can now be
rewritten as
\begin{align}
  H =& \left( -
       J_1 \sum\limits_{\Braket{ij}} c^\dag_{\vec{r}_i}
       e^{i\left(\alpha_{\vec{r}_i}-\alpha_{\vec{r}_j}\right)} c_{\vec{r}_j} +
       J_2 \sum\limits_{\Braket{\Braket{ik}}} c^\dag_{\vec{r}_i}
       e^{i\left(\alpha_{\vec{r}_i}-\alpha_{\vec{r}_k}\right)}
       c_{\vec{r}_k} + \text{h.c.} \right)
       \;.
\end{align}

\begin{figure}
  \centering
  \includegraphics[width=0.85\textwidth]{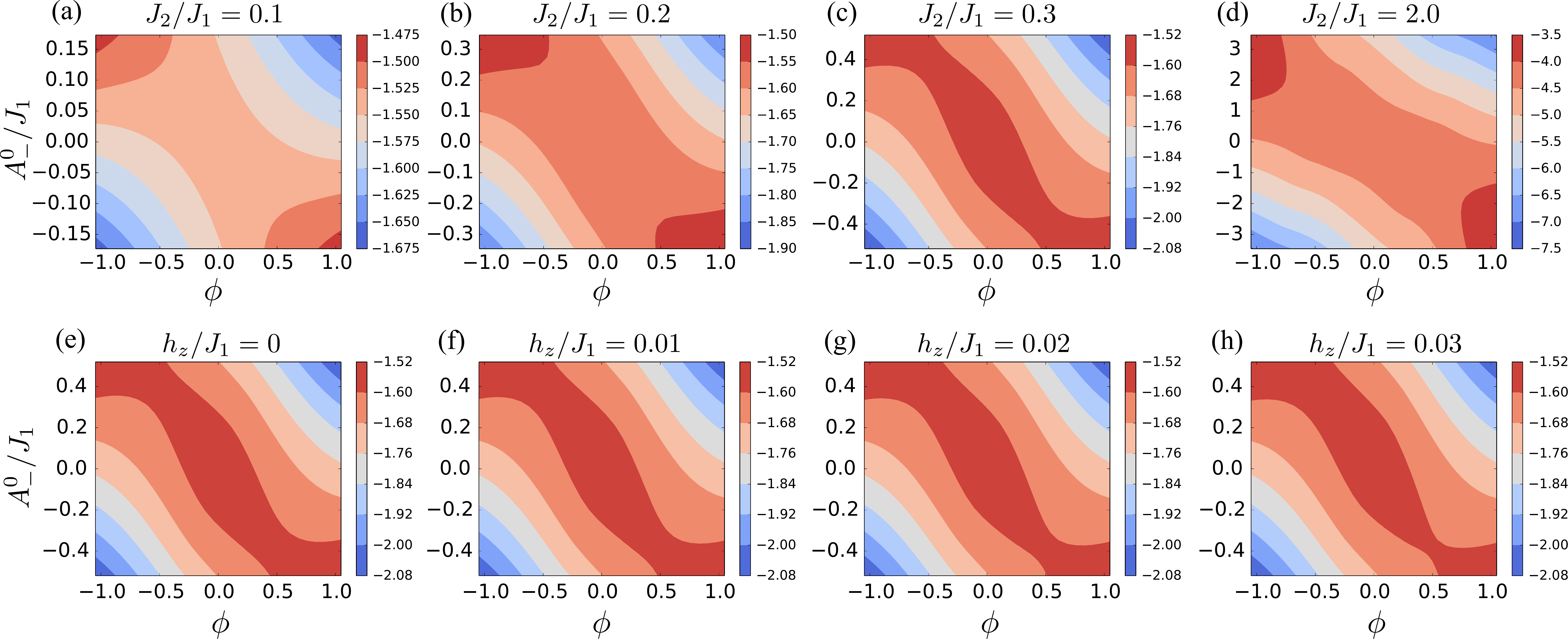}
  \caption{\textbf{(a-d)} The functional
    $\mathcal{L}_\text{eff}(\phi, A^0_-)$ of Eq.~\eqref{eq:Leff}
    plotted in units of $J_1$ for different values of $J_2/J_1$,
    $h_z = 0$. \textbf{(e-h)} The effect of the $\mc{P}$ breaking term
    $H_z$ on the functional $\mathcal{L}(\phi, A^0_-)$ for a fixed
    value $J_2 / J_1 = 0.3$. We can see that one of the non-trivial
    minima shifts in energy with respect to another one, explicitly
    breaking the symmetry between two degenerate solution from the
    $h_z = 0$ case.}\label{fig:6}
\end{figure}

We introduce a vector field $\bm{A} \left( \vec{r}_k \right)$ defined
as
\begin{equation}
  \Braket{\alpha_{\vec{r}_j} - \alpha_{\vec{r}_i}} =
  \int\limits_{\bm{r}_i}^{\bm{r}_j} \textrm{d}\bm{r}_k
  \cdot \bm{A} \left( \vec{r}_k \right) \;,
\end{equation}
and a ChS magnetic field
$\bm{B} \left( \vec{r}_i \right) = B \left( \vec{r}_i \right)
\vec{e}_z$ such that
\begin{equation}
  B \left( \vec{r}_i \right) = \text{curl} \bm{A} \left( \vec{r}_i
  \right) = 2 \pi \Braket{n_{\vec{r}_i}} = 2 \pi n(\vec{r}_i) \;.
\end{equation}
We remove exponential string operators by introducing the
$\delta$-function imposing a constraint on the ChS magnetic field
through the Lagrange multiplier $A^0(\bm{r}_i)$:
\begin{equation}
  2\pi \delta\left(B(\bm{r}_i)/(2\pi) - n({\vec{r}_i})\right) =
  \int { \textrm{d} A^0(\bm{r}_i)
    \exp \left\lbrace i A^0(\bm{r}_i) \left[ B(\bm{r}_i)/(2\pi) -
        n({\vec{r}_i}) \right] \right\rbrace }
  \;.
\end{equation}
We write down the resulting action
\begin{align}
  S = \int \textrm{d}t
  & \left[
    \sum\limits_i \bar{\psi}(\bm{r}_i) \left( i\partial_t -
    A^0(\bm{r}_i) \right) \psi(\bm{r}_i) +
    \frac{1}{2\pi} \sum\limits_i A^0(\bm{r}_i) B(\bm{r}_i)
    \right.
    \notag \\
  & \left. -
    J_1 \sum\limits_{\Braket{ij}} \bar{\psi}(\bm{r}_i) \psi(\bm{r}_j)
    e^{i\left\langle\alpha_{\vec{r}_i}-\alpha_{\vec{r}_j}\right\rangle} +
    J_2 \sum\limits_{\Braket{\Braket{ik}}}
    \bar{\psi}(\bm{r}_i) \psi(\bm{r}_k)
    e^{i\left\langle\alpha_{\vec{r}_i}-\alpha_{\vec{r}_k}\right\rangle}
    + \text{h.c.}
    \right]
    \;.
\end{align}
The functional integration with respect to the ChS magnetic field
$B(\bm{r}_i)$, the Lagrange multiplier $A^0(\bm{r}_i)$ playing the
role of the scalar potential and Grassman variables
$\bar{\psi}(\bm{r}_i)$ and $\psi(\bm{r}_i)$ associated to fermionic
creation and annihilation operators is considered. One can integrate
out Grassmann variables. At the mean-field level we express the
fermionic free energy functional
$W(\lbrace A^0(\bm{r}_i), B(\bm{r}_i) \rbrace)$ as a sum over
eigenvalues of the single-particle problem up to the Fermi energy in
such a way that the total filling of fermions equals $1/2$:
\begin{align}
  W(\lbrace A^0(\bm{r}_i), B(\bm{r}_i) \rbrace)
  =& \sum\limits_k
     E_k(\lbrace A^0(\bm{r}_i), B(\bm{r}_i) \rbrace)
     \Theta( E_k - E_F)
     \;, \notag \\
  N_c
  =& \sum\limits_k \Theta( E_k - E_F)                                
  \;.
\end{align}
Here $N_c$ is the total number of unit cells in the lattice, $\Theta$
is the Heaviside function and $E_F$ is the Fermi energy, that is
calculated self-consistently. We suppose that the solution is
translation invariant. In particular, $n({\vec{r}_i}) = n_A$ or $n_B$.
We allow however the breaking of the symmetry between two sublattices:
$n_A \neq n_B$. The condition of being at total filling $1/2$ implies
$n_A + n_B = 1$. The first-neighbor hopping terms are sensitive only
to the total flux through the unit cell $\Phi_\text{Tot} = 2 \pi$
(each unit cell containing precisely 1 site of the sublattice $A$ and
1 site of the sublattice $B$), that is gauge equivalent to
zero. Second-neighbor hoppings exhibit Haldane modulations of the flux
through big triangles formed by second-neighbor links. In order to see
this more clearly, we separate a symmetric $(+)$ and an antisymmetric
$(-)$ components of the scalar potential and the magnetic field:
$ B_\pm = B_A \pm B_B, \ A^0_\pm = A^0_A \pm A^0_B $.  The flux
configuration due to the symmetric component is also gauge equivalent
to zero for second-neighbor links, whereas the antisymmetric component
leads to $\Phi_A = - \Phi_B = B_{-}$. Here $\Phi_A$ and $\Phi_B$ are
fluxes through the smallest triangles formed by second-neighbor sites
of the sublattice $A$ or $B$. For consistency with the notation of the
Ref.~\onlinecite{SedrakyanGlazmanKamenev2015}, we also define
$\phi = B_- / 3$.  The resulting effective Lagrangian for the ChS
magnetic field and the scalar potential is
\begin{equation}
  \label{eq:Leff}
  \mathcal{L}_\text{eff}( A^0_-, \phi ) = W( A^0_-, \phi ) +
  \frac{3N_c}{2 \pi} A^0_- \phi
  \;.
\end{equation}
The effective mean-field model for free fermions is the Haldane
model~\cite{Haldane1988}. We use the saddle-point approximation to
find the values of $A^0_-$ and $\phi$:
\begin{equation}
  \delta_{A^0_-} S_\text{eff} = 0,\quad \delta_{\phi} S_\text{eff} = 0
  \;.
\end{equation}
Solutions of these equations correspond to the extrema of the
functional $\mathcal{L}_\text{eff}$, as shown in Fig.~\ref{fig:6}. By
calculating this functional for different values of $J_2/ J_1$, we
deduce three different regimes in the phase diagram. In the region
$J_2 / J_1 \lesssim 0.21$ the functional $\mathcal{L}_\text{eff}$ has
only one point where both equations are verified, that is the saddle
point at $A^0_- = 0$, $\phi = 0$.  In the region
$0.21 \lesssim J_2 / J_1 \lesssim 0.36$ there are three solutions of
the equations for the minimization. The solution at $A^0_- = 0$,
$\phi = 0$ corresponds to a local maximum of the functional
$\mathcal{L}_\text{eff}$, whereas two symmetric solutions not located
at zero become new saddle point solutions. These solutions moves
continuously with $J_2/J_1$, starting from zero, that corresponds to a
second order phase transition. In the region $0.36 \lesssim J_2 / J_1$
again only the local minimum of $\mathcal{L}_\text{eff}$ remains as a
solution at $A^0_- = 0$, $\phi = 0$, that corresponds to a first order
phase transition.

\subsection{Effect of the external magnetic field $h_z$}

We consider the effect of adding an external magnetic field $h_z$ to
the mean-field solution. In the expression of the fermionic
single-particle spectrum this term appears as a Semenoff mass
term~\cite{Semenoff1984}. By doing the numerical minimization, we see
that the effect of this perturbation consists in breaking the symmetry
between two non-trivial solutions in the regime
$0.2 \lesssim J_2 / J_1 \lesssim 0.36$. This effect is presented in
Fig.~\ref{fig:6}.

\section{Exact diagonalization: Classical phases of the frustrated
  spin-1/2 $XY$ model}

\begin{figure}
  \includegraphics[width=0.48\textwidth] {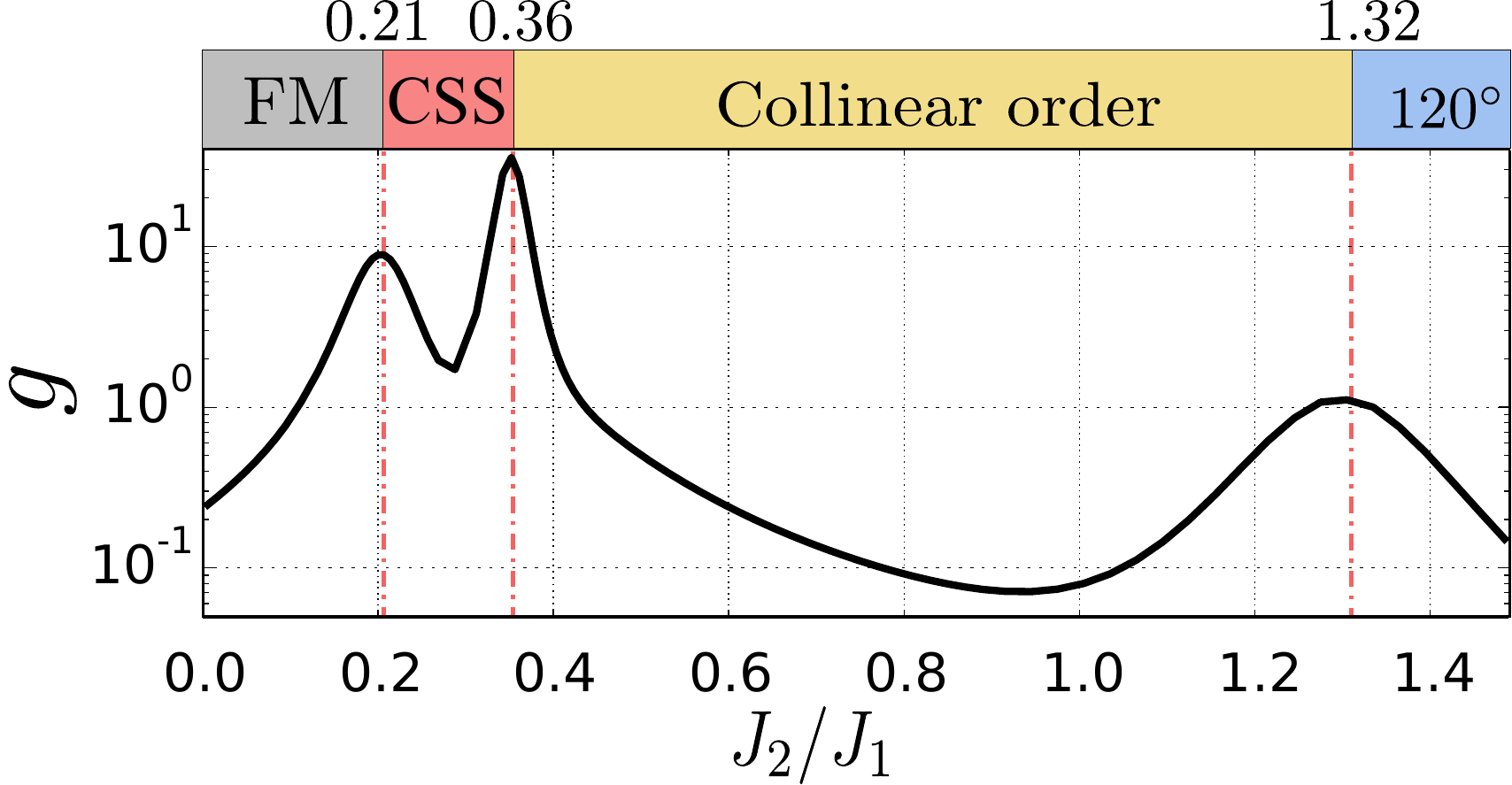}
  \caption{ED calculation of the fidelity metric $g$ on a lattice of
    $4 \times 3$ unit cells for hard-core bosons at filling $1/2$
    ($S^z_\text{Tot} = 0$).}
  \label{fig:4}
\end{figure}

\begin{figure}
  \includegraphics[width=0.95\textwidth] {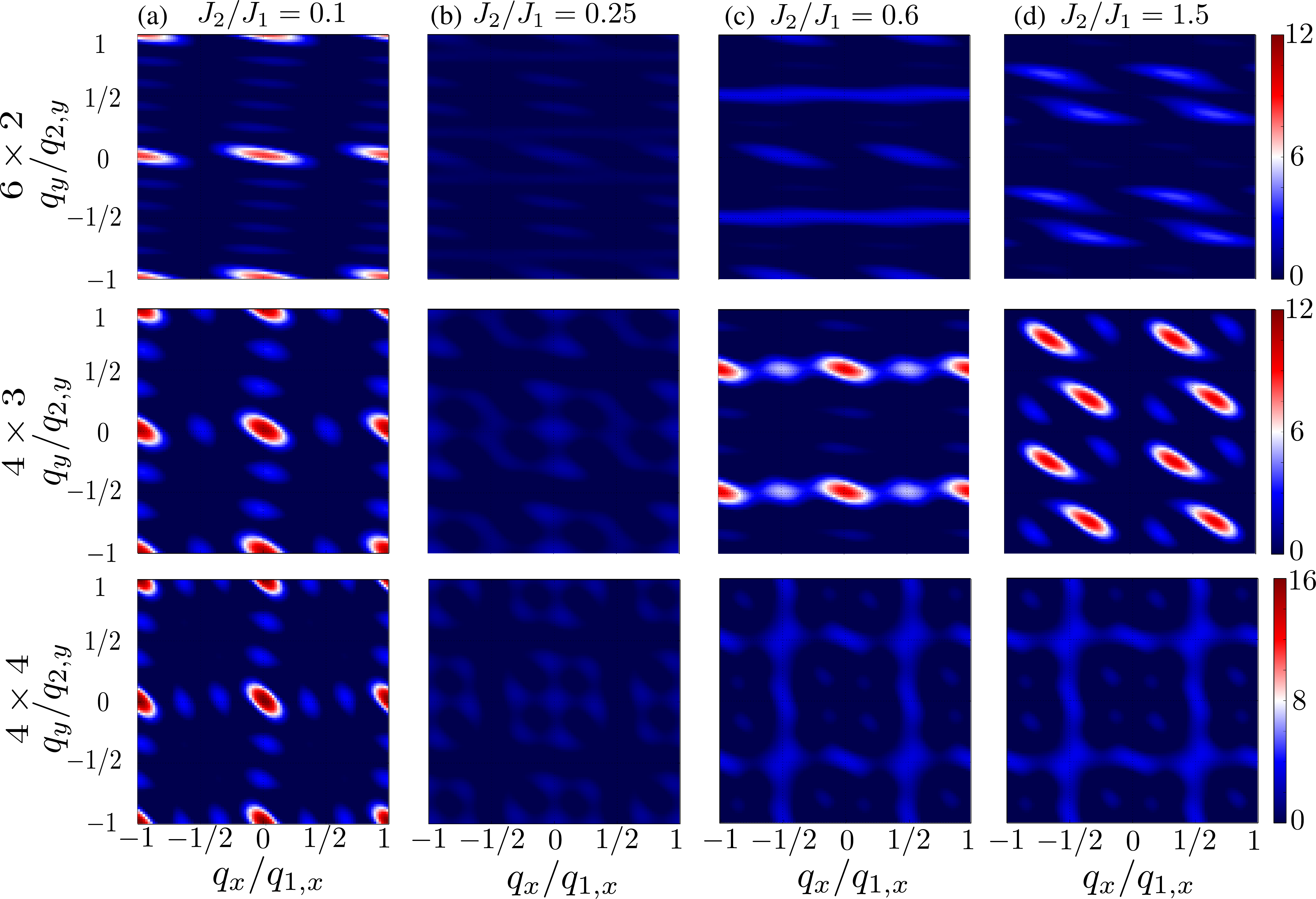}
  \caption{ED calculation of the static structure factor
    $S_\textrm{Spiral}\left( \vec{q} \right)$ at 4 typical points in 4
    phases (different rows) on various lattices (different lines) for
    hard-core bosons at filling $1/2$ ($S^z_\text{Tot} = 0$). Vectors
    $\vec{q}_1$ and $\vec{q}_2$ are defined as in Fig.~\ref{fig:0}(a).
    \textbf{(a)} In the FM phase ($J_2 / J_1 = 0.1$ row) the structure
    factor is piked at $\vec{q} = \vec{\Gamma}$. \textbf{(b)} The
    systems seems to be disordered in the $xy$ plane in the
    intermediate frustration regime ($J_2 / J_1 = 0.25$
    row). \textbf{(c)} We observe a formation of the collinear order
    for $J_2 / J_1 = 0.6$. We notice however the significant
    difference of the result on the lattice $4 \times 3$. This is
    explained by the fact that this lattice does not contain all the
    $\vec{M}$ points in the reciprocal space. \textbf{(d)} In the case
    $J_2 / J_1 = 1.5$ the system forms a $120^\circ$ order. We notice
    that similarly to the previous case the lattice $4 \times 4$ does
    not contain Dirac points $\vec{K}$ in the reciprocal space, that
    results in the impossibility to recover correctly the $120^\circ$
    phase: two rightmost figures in the bottom line do not differ
    almost at all.}
  \label{fig:5}
\end{figure}

In order to determine the phase boundaries of the frustrated spin-1/2
$XY$ model, we calculate the fidelity metric
$g$~\cite{ZanardiPaunkovic2006, ShiJian2010, VarneyEtAl2010}. The
result of this calculation on the lattice of $4 \times 3$ unit cells
is shown in Fig.~\ref{fig:4}.
Classical phases are studied by looking at the correlation functions
$\Braket{S_{\vec{r}_i}^\mu S_{\vec{r}_j}^\nu}$ and the related
coplanar structure factor
\begin{equation}
  S_\textrm{Spiral}\left( \vec{q} \right) = 2 \sum\limits_{i, j \in A}
  e^{i\vec{q} \cdot \left( \vec{r}_i - \vec{r}_j \right)}
  \braket{S^x_{\vec{r}_i} S^x_{\vec{r}_j}} \;.
\end{equation}
The result of such analysis is presented in Fig.~\ref{fig:5}.

\end{document}